\documentclass[fleqn,10pt,twocolumn]{wlscirep}

\usepackage{caption}
\usepackage{subcaption}
\usepackage{picins}

\title{Everything You Wanted to Know About Space Radiation but Were Afraid to Ask}

\author[1,2,9,*]{Chancellor, JC}
\author[3]{Nowadly, C}
\author[6]{Williams, JA}
\author[2,7,8]{Aunon-Chancellor, SM}
\author[1]{Chesal, ME}
\author[4]{Looper, J}
\author[1,5]{Newhauser, W}
\affil[1]{Department of Physics \& Astronomy, Louisiana State University, Baton Rouge, LA, USA}
\affil[2]{Department of Preventive Medicine \& Population Health, University of Texas Medical Branch, Galveston, TX, USA}
\affil[3]{Department of Emergency Medicine, Brooke Army Medical Center, Fort Sam Houston, TX}
\affil[4]{Department of Veterinary Clinical Sciences, LSU School of Veterinary Medicine, Baton Rouge, LA, USA}
\affil[5]{Department of Physics, Mary Bird Perkins Cancer Center, Baton Rouge, LA, USA}
\affil[6]{Departments of Environmental Medicine \& Radiation Oncology, University of Rochester Medical Center, Rochester, USA}
\affil[7]{Department of Internal Medicine, LSU Health Science Center, Baton Rouge, LA}
\affil[8]{Astronaut Office, NASA Johnson Space Center, Houston, TX}
\affil[9] {Fellow, Outerspace Institute, University of British Columbia, CA}
\affil[*]{Author to whom correspondence should be addressed; jeff@spartanphysics.com}

\keywords{space radiation; radiobiology; aerospace medicine}
\begin{abstract}
	\textbf{The space radiation environment is a complex combination of fast-moving ions derived from all atomic species found in the periodic table. The energy spectrum of each ion species varies widely but is prominently in the range of 400–600 MeV/n. The large dynamic range in ion energy is difficult to simulate in ground-based radiobiology experiments. Most ground-based irradiations with mono-energetic beams of a single one ion species are delivered at comparatively high dose rates. In some cases, sequences of such beams are delivered with various ion species and energies to crudely approximate the complex space radiation environment. This approximation may cause profound experimental bias in processes such as biologic repair of radiation damage, which are known to have strong temporal dependancies. It is possible that this experimental bias leads to an overprediction of risks of radiation effects that have not been observed in the astronaut cohort. None of the primary health risks presumely attributed to space radiation exposure, such as radiation carciogenesis, cardiovascular disease, cognitive deficits, etc., have been observed in astronaut or cosmonaut crews. This fundamentally and profoundly limits our understanding of the effects of GCR on humans and limits the development of effective radiation countermeasures.}
\end{abstract}
\begin{document}

\flushbottom
\maketitle \thispagestyle{empty}

\section*{Why do we need yet another review of space radiation research?} Space travelers are exposed to a myriad of environmental stressors, including chemicals from equipment,“microbes” from occupants of the space vehicle, and microgravity, to name a few. Cosmic radiation exposures are a particularly challenging environmental stressor because they vary strongly in magnitude and quality with vehicle trajectory, mission duration, and solar activity. Rare but potentially-lethal solar eruptions of protons and other particles, which could result in mission failures, have so far proven utterly unpredictable, precluding mission timing as a mitigation strategy \cite{carnell2016risk,chancellor2018limitations}. To complicate matters further, the risks to humans vary strongly with the radiation exposure and the host, e.g., age, sex, and genetic profile of each individual. Risk projections for individuals astronauts are highly uncertain and controversial. Although some progress has been made toward understanding and mitigating the risks of radiation to astronauts, the overall situation has scarcely improved since the dawn of crewed spaceflight \cite{chancellor2018limitations,chancellor2014space}.

Several approaches are possible to mitigate space radiation risks. The use of shielding can reduce the radiation exposures. Passive shields use bulk material fields , e.g., the vehicle, habitat and or local regolith, to attenuate radiation, whereas proposed magnet shields deflect charged- particle radiation away from habituated areas \cite{cucinotta2006evaluating,guetersloh2006polyethylene,newhausershielding,wilson1999shielding,zeitlin2005shielding,slaba2017optimal}. The expecteds exposures can be reduced by shortening the mission, e.g., with higher-impulse propulsion systems such as ion propulsion drives. Other approaches include radiation sensitivity as crew selection criterion, biologic countermeasures (e.g., radioprotective drugs), increased medical surveillance of exposed crew, and even pre-exposure prophylactic surgical removal of the sensitive female breast has been proposed. More than a half century later, intensive research has not revealed a method to fully mitigate the radiation risk, and our understanding of the physics and biology of these risks remains incomplete. Consequently, radiation risk prediction and management leads as the major unmet challenge for planning future crewed missions, especially as spaceflight missions increase and we travel outside the earth’s protective magnetic field \cite{chancellor2018limitations}.

A promising approach to solving major gaps our the knowledge on radiation effects is to rely on ground based research involving those effects in animals. Research efforts have been frustrated by a scarcity of robust biologic and physical surrogates for such experiments. These obstacles have made it difficult to translate radiation risk models from animal models to humans \cite{williams2010animal,williams2016addressing}. The health risks to spaceflight explorers because of the space radiation environment, therefore remains incomplete, compounded by the disparities between findings from space-based observations of human astronauts vs. ground-based experiments. Recently, we proposed new methods and avenues of inquiry that overcome many of the obstacles previously encountered \cite{chancellor2017targeted}. Here we will discuss several methodological aspects of ground-based experimentation which, if improved, will more faithfully mimic the complex radiation environment encountered in a variety of spaceflight missions.

\begin{figure}[ht]
\centering
\includegraphics[width=\linewidth,height=7cm,keepaspectratio]{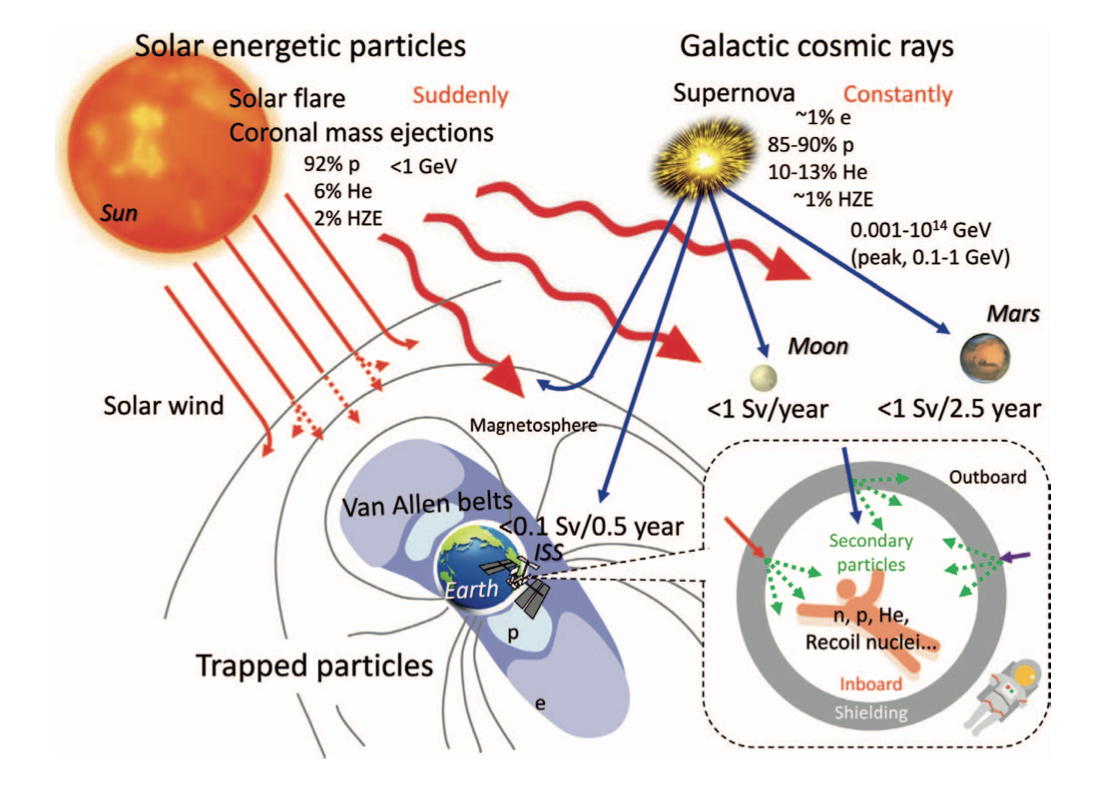}
\caption[]{{\small{\bf Components of the space radiation environment.} Operational space radiation environment is comprised of three sources of ionizing radiation, energetic protons from solar particle events, relativistic heavy ions of the galactic cosmic ray spectrum, and trapped electrons and protons in the Van Allen Belts.
Image courtesy of ESA.}}
\label{fig:fig1}
\end{figure}

\section*{What exactly is the operational space radiation environment?}
\begin{figure}[ht]
  \centering
  \includegraphics[scale=0.6]{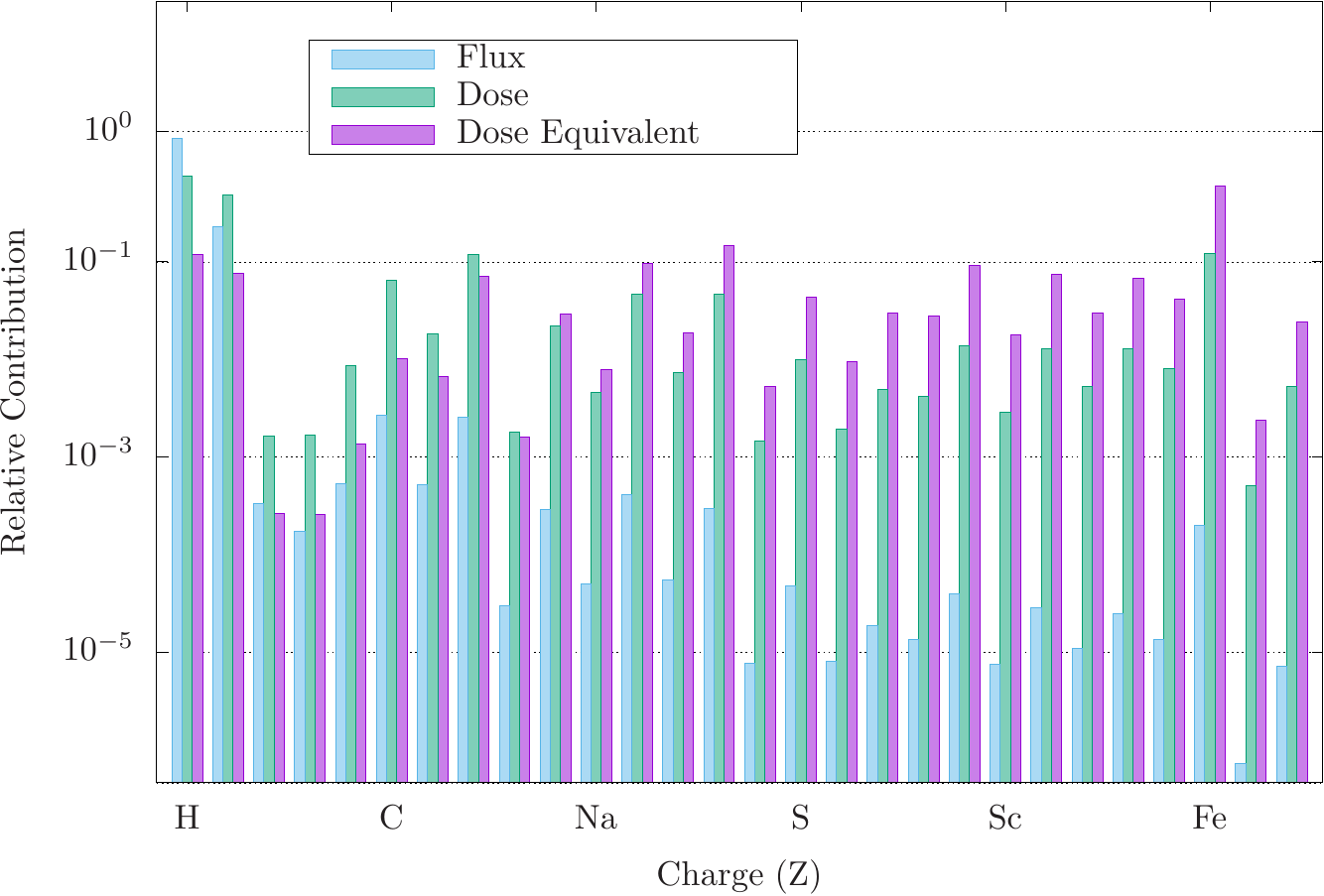}
  \caption[]{{{\small \bf Relative abundance of the GCR spectrum, and the contribution to dose and dose equivalent from hydrogen (Z = 1) to nickel (Z = 28).}  Heavier ions such as Z = 26 make up a relatively small portion of the spectrum, however, they contribute larger energy doses to the cumulative radiation exposure than even hydrogen or helium ions.}}
  \label{fig:fig2}
\end{figure}

The operational space radiation environment, shown in Fig.~\ref{fig:fig1}, can be divided into three separate ionizing radiation sources, solar wind consisting of mostly low energy protons and electrons, heavy-charged particles found in the Galactic Cosmic Ray (GCR) spectrum, and energetic protons associated with a Solar Particle Event (SPE)\cite{chancellor2018limitations}. The background dose rate for solar-wind protons varies with the solar cycle (9-14 year period, average of 11 years). Even at solar maximum, the contribution is much less than that from GCR and therefore is considered a negligible risk\cite{chancellor2014space}. GCR nuclei originate from outside our solar system and possess sufficient energies to penetrate any shielding used on current mission vehicles. As demonstrated in Fig.~\ref{fig:fig2}, the GCR spectrum consists of about 87\% hydrogen ions (Z=1 or protons) and 12\% helium ions (Z=2 or alpha particles), with the remaining 1-2\% heavier nuclei with charges ranging from Z=3 (Lithium) to about Z=28 (Nickel)\cite{chancellor2018limitations,simpson1983elemental}. Ions heavier than nickel are also present, but they are rare in occurrence. The fluence (the number of incident particles crossing a plane of unit area) of GCR particles in interplanetary space range fluctuates inversely with the solar cycle, with dose rates of 50 to 100 mGy/year at solar maximum to 150 to 300 mGy/year at solar minimum\cite{mewaldt2005cosmic,rahmanifard2020galactic,wouter2020crater}, During spaceflight transit outside of low-Earth Orbit (LEO), every cell nucleus within an astronaut would be traversed by a hydrogen ion or energetic electron (e.g., delta ray) every few days, and by the heavier GCR ion (e.g., O, Si, and Fe ions) every few months\cite{chancellor2014space}. Heavy ions, despite their comparatively lower fluences, contribute a significant amount to the GCR dose that astronauts will incur outside of LEO because of their large ionization power because of their large ionization power. Shown in Fig.~\ref{fig:fig2} is the GCR flux, dose, and dose equivalent\footnote{Dose equivalent, $H$, is a measure of the biological damage to living tissue as a result of radiation exposure and is calculated as the product of absorbed dose, $D$, in tissue multiplied by a quality factor, $Q$, the is dependant on the linear energy transfer of the radiation particulate, $H = QD$. The SI units of dose equivalent is sieverts (Sv).} up to nickel, Z=28. Light ion species such as hydrogen and helium make up most of the GCR spectrum, but heavier ions such as silicon (Z=14), iron (Z=26), etc., contribute significantly once the biological equivalent dose is factored. The swiftest of the GCR heavy ions are so penetrating that shielding can only partially reduce the intra-vehicular (IVA) doses\cite{cucinotta2006cancer}. In theory, more massive shielding could provide some additional protection, but in practice this is limited by the payload lift capabilities of spacecraft launch systems. In fact, studies have shown that aluminum shielding equivalent to 16 times the thickness of the Apollo command module can reduce the GCR exposure by up to 25\%. Similarly, the equivalent mass of polyethylene, a better shielding material but an inferior structural material, would only provide a \%35 reduction in GCR dose\cite{board1997radiation,edwards2001rbe}.

SPEs consist of high-energy protons that emanate from the Sun in regions of magnetic instability\cite{ncrp2006information}. SPE radiation is primarily composed of protons with kinetic energies ranging from 10 MeV up to several GeV (Giga electron Volts). The fluence from and occurrence of SPEs is unpredictable, but dose-rates as high as 1500 mGy/hour (approximately 3000mSv/hour)\footnote{It should be noted that the nominal dose-rate onbard the International Space Station (ISS) is approximately 0.04 mSv/hour and predicted to be around 0.13 mS/hour during space travel outside LEO.} have been measured\cite{ncrp1988guidance}. During an SPE, a localized magnetic disturbance of the Sun resultss in the release of intense bursts of ionizing radiation that follow magnetic field lines. Their irregular and unpredictable occurrence is evidenced by the unusual occurrence of four comparably large SPEs in a period of 4 months during the 22nd solar cycle\cite{wilson1999shielding,hellweg2007getting,smart2003comment}. Computer simulations reveal that the dose distribution in an astronaut would be highly non-homogeneous, with a relatively high superficial dose and a lower internal dose. Extravehicular exposures carry higher risks than intravehicular exposures because the shielding provided by the space suit is much less than that of the vehicle. Even inside the vehicle, an SPEwould still expose astronauts within the spacecraft to non-negligible levels of radiation. Although large SPE doses can instigate so-called late-occuring effects such as radiogenic cancer, ocular cataracts, respiratory and digestive diseases, and damage to the microvasculature, these are mostly latent for many months, years, or even decades after exposure, and by definition they should not manifest as an immediate risk to crew health during the mission\cite{chancellor2014space}.

As future missions are planned outside LEO and away from the protection of the Earth’s magnetic shielding, the range of possible radiation exposures that astronauts may encounter is markedly different from those of LEO. The risks from GCR and SPE exposures are increased. As previously mentioned, massive shielding in spacecraft can reduce but not eliminate these exposures. In fact, collisions between GCR particles and the nuclei of shielding materials can initiate an avalance (sometimes called a nuclear cascade or shower) of nuclear reactions. Consequently, niaively increasing the shielding mass can actually increase the radiation risk from secondary particles from the avalanches. The optimal mass of shielding to attenuate SPE radiation is thought to be 30 to 40 g/cm$^{2}$\cite{chancellor2018limitations,chancellor2017targeted}. This amount of radiation shielding is somewhat impractical given lift capabilities in current rockets.

\section*{What analogs are used to evaluate the space radiation risk to human health?}

Animal models are used as human surrogates in studies of radiobiology to obtain data that cannot be gained in ethical studies of humans. Many animal models have been developed to explore a wide variety of questions in space radiation protection and medicine, e.g., radiation oncology. Mammalian species and other advanced organisms, especially those that can reproduce quickly and possess genomes similar to humans, have the utility to identify mechanisms for radiation-induced effects and diseases \cite{williams2010animal,nelson2016space}. These experiments have contributed significantly to our understanding of disease mechanisms, but their relevance to and suitability for predicting outcomes in humans is controversial, e.g., for non-astronaut patients \cite{hackam2006translation,perel2007comparison} and astronauts alike \cite{chancellor2018limitations,kennedy2014biological}. A major limitation of current animal models is that animals are, in some relevant ways, profoundly different from humans in their response to radiation, e.g., at the cellular, organ, and organism levels. This could be explained by limitations in current methods in animal studies, leading to systematic bias, flawed data, and and uncertainty of the validity of qualitative conclusions regarding efficacy of, for example, countermeasures \cite{chancellor2018limitations,williams2010animal}. Rodent models may be less well suited to test outcomes in normal tissue responses to radiation in humans due to their high level of genetic instability. There is evidence that larger mammalian models such as canines provide a better model for radiation research. DNA repair mechanisms are highly conserved between mammalian species, and that there is high homology between key DNA damage response genes in humans and dogs \cite{williams2010animal,nolan2019emerging}. Numerous studies in dogs have modeled normal tissue radiation response and those studies have helped optimize human oncology care, however, no large studies of canine models have been used in space radiation studies on normal tissue effects.

Limitations in the ability to translate study outcomes into equivalent health consequences in humans also result from critical disparities, usually disease-specific, between the animal models and the human subject in clinical trials and  space missions, for example the development of harderian gland tumors which only appear in mice \cite{williams2010animal,williams2016addressing,siranart2016mixed}. Additionally, because non-human mammals are genetically different from humans, they may yield results that are difficult to interpret. A growing list of genes is known to affect radiation sensitivity phenotypes for numerous radiation effects, such as molecular, chromosomal, signal transduction associated growth-regulating changes, cell killing, tumor acute and late effects, and animal carcinogens. To date, however, there is still no knowledge that would allow a direct connection to be drawn between observed changes in gene sequence and a corresponding change in radiosensitivity. Recently, however, promising results from Edmondson et al, demonstrated in a mouse model that the underlying genetics of susceptibility can be similar for tumorigenesis following exposure to both high- and low-LET radiation. This indicates that epidemiology studies from human exposures to gamma radiation may be of utility for predicting the cancer risks attributed to GCR, but further work is needed to validate these findings \cite{edmondson2020genomic}.

\section*{Exactly how do you estimate radiogenic risk?}
It is believed that the high \emph{linear energy tranfer} (LET) radiation in the GCR spectrum can induce cancer, cognitive deficits, changes associated with premature aging and degenerative effects in many organs. The LET quantifies how much energy is lost by an ion, on average, per unit pathlength traversed in matter, is typically given in units of kilo electron volts per micron (keV/$\mu$m), and depends strongly on ion velocity for quantification of radiobiological damage. Additionally, the large ionization density of GCR ions makes them a potentially significant contributor to cellular and organ damage \cite{chancellor2014space,walker2013heavy}, and therefore, it is essential to study the potential health risks from GCR exposures. There are numerous limitations of current terrestrial analogs used to study and predict the effects of space radiation on biologic systems. The mechanisms that cause biological damage from GCR differ from those from traditional terrestrial radiation sources. Terrestrial analogs often use radiation that causes indirect ionizing events \cite{chancellor2018limitations,blue2019limitations}.

The biological effect of the radiation dose depends on physical and biological factors, e.g., multiple particle and energy-specific factors, dose rate per exposure and the frequency of multiple exposures.  The (physical) absorbed dose is the energy absorbed per mass (J/Kg, Gy).  For a dose-based system of radiation protection and for the determination of occupational dose limits, it is necessary to attempt summing the total risk of radiation from multiple sources (e.g., SPE protons, GCR, etc.) \cite{hall2006radiobiology}.  At a given ion velocity,  LET increases with atomic number. Thus, for ground-based research, it is key to have the correct abundances and energy distributions of each ion present in the space radiation environment.  As charged particles lose energy successively through material interactions, each energy loss event can result in damage to the biological tissue. In addition, as charged particles near the end of their track (i.e., as they slow down and are nearly stopped) the LET rises sharply, creating the so-called “Bragg peak” \cite{icrp1998icrp}.
\begin{figure}[ht]
	\centering
	\includegraphics[scale=0.60]{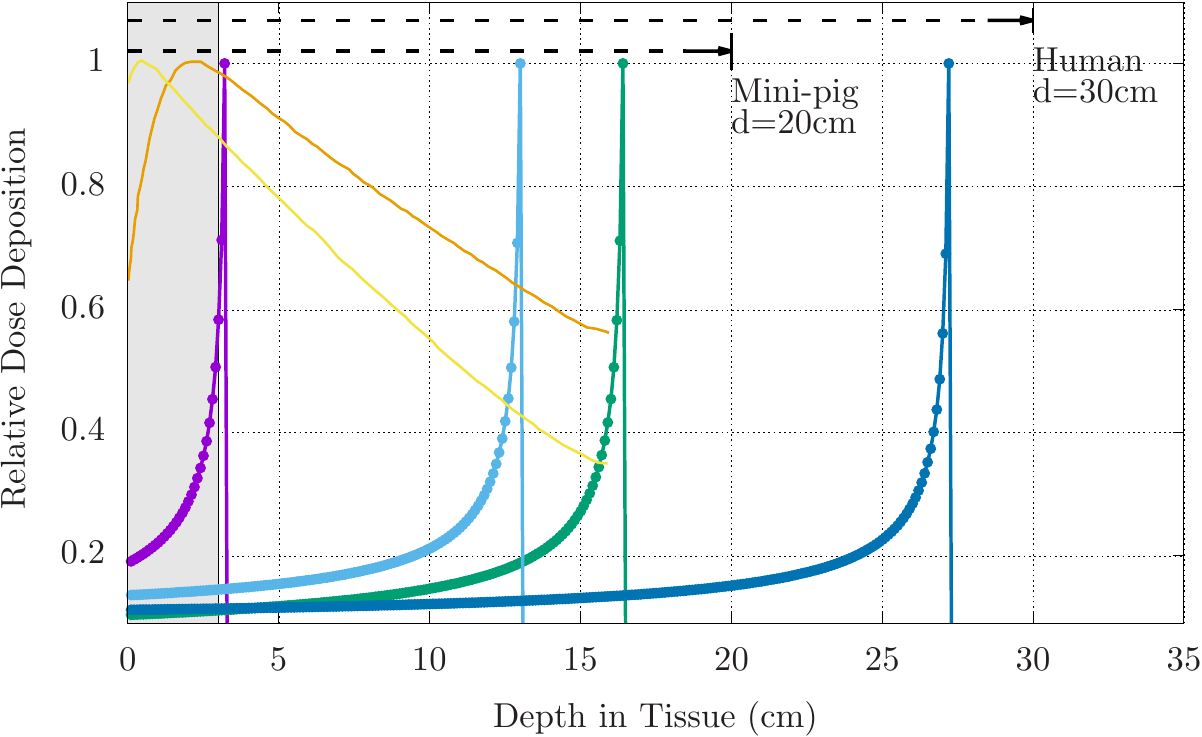}
	\caption{{\small{\bf Bragg peak and depth dose characteristics of space radiation.}   Shown in the figure are the calculated Bragg peaks and relative dose deposition for ions at energies commonly used in space radiation studies. These are compared to the x-ray and gamma sources used as surrogate radiations for RBE quantification. This effect is very pronounced for fast moving, charged particles. Shown are 60 MeV protons (hydrogen, purple), 600 MeV/n 56Fe (iron, light blue), 290 MeV/n 12C (carbon, green), 1 GeV/n 56Fe (iron, dark blue), x-ray (orange dotted line), and 60Co (cobalt, yellow dotted line). The shaded gray area, representing the average diameter of a mouse, demonstrates that the Bragg peak, and thus the majority of dose deposition, is outside the mouse body for SPE protons (energies 50 MeV/n) and GCR ions. Figure reprinted with permission from Chancellor et al. \cite{chancellor2018limitations} under the Creative Commons license.}}
	\label{fig:fig3}
\end{figure}
The phenomena of the Bragg peak is exploited in cancer therapy in order to concentrate the dose at the target tumor while minimizing impact to the surrounding tissue. This is demonstrated in Fig.~\ref{fig:fig3} where the relative dose deposition in tissue for various radiation types utilized in space radiobiology studies plotted versus depth in tissue. The gray shaded area is the average width of a mouse model. Also shown are the average diameters of Yucatan mini-pigs and humans. Gamma and X-ray radiations deposit most of the energy at or near the surface, while in contrast, charged particles such as protons, carbon, iron, etc., have distinct Bragg peaks. In each example, the Bragg peak is located outside the body mass of the mouse, indicating the difficulty in replicating the relative organ dose distribution of a GCR exposure incurred by humans during spaceflight.

The \emph{relative biological effectiveness} (RBE)\footnote{The RBE is defined as the ratio of doses between some test radiation, D$_{r}$, and the dose of x-rays, D$_{xray}$, required for equal biological effect, RBE = D$_{xray}$/D$_{r}$ \cite{hall2006radiobiology}.} quantifies how much energy is required for a test radiation (e.g., 1-GeV protons) compared to a reference radiation (e.g., 140 KeV x-rays).  RBE depends on ion charge and speed, the cell, organ, or tissue of interest, amd the endpoint considered, (e.g., lethal damage and carcinogenesis).  To facilitate the aggregation of risk to a human, variations in radiation sensitivities in various tissues and organs are incorporated in a tissue weighting factor.  Similarly, variations in radiation sensitivity with ion properties are taken into account with radiation weighting factors, which are informed by RBE values (mostly for the endpoint of carcinogenesis).  Additional factors may be applied to take into account variations in dose, dose rate, and other factors.  Combining the physical absorbed dose and the various biologic considerations mentioned, one obtains an “equivalent dose” in units of Sieverts (Sv) that ideally is proportional to risk in the subject or population being considered\cite{hall2006radiobiology,attix2008introduction}.

In the US, radiation therapy with charged particles is dominated by protons.  An RBE value for proton beams was recommended at 1.1 for all endpoints, tissues, doses, LET values, and dose rates. This consensus recommendation was made mainly to increase consistency in dose prescription and reporting among proton therapy centers.  It also is believed to facilitate pooling of data from observational data from photon beam treatments.  Importantly, to date, there remains insufficient evidence that the RBE is significantly different from 1.0 for any human tissue for any endpoint. The ICRU [report 78] and the 2019 report from the American Association of Physicist in Medicine described limitations in RBE data for clinical purposes. Based on the lack of evidence outlined above, we can conclude that many of the same gaps in knowledge comprise obtacles to reliable risk predictions in both radiation oncology and crewed spaceflight \cite{icrp1998icrp,dasu2013impact}.  This suggests  the possibility of synergistic research studies of open questions in oncology and space exploration.

\section*{What outcomes have been seen in astronauts so far?}
\begin{figure*}[ht]
\centering
\includegraphics[width=\linewidth,height=8cm,keepaspectratio]{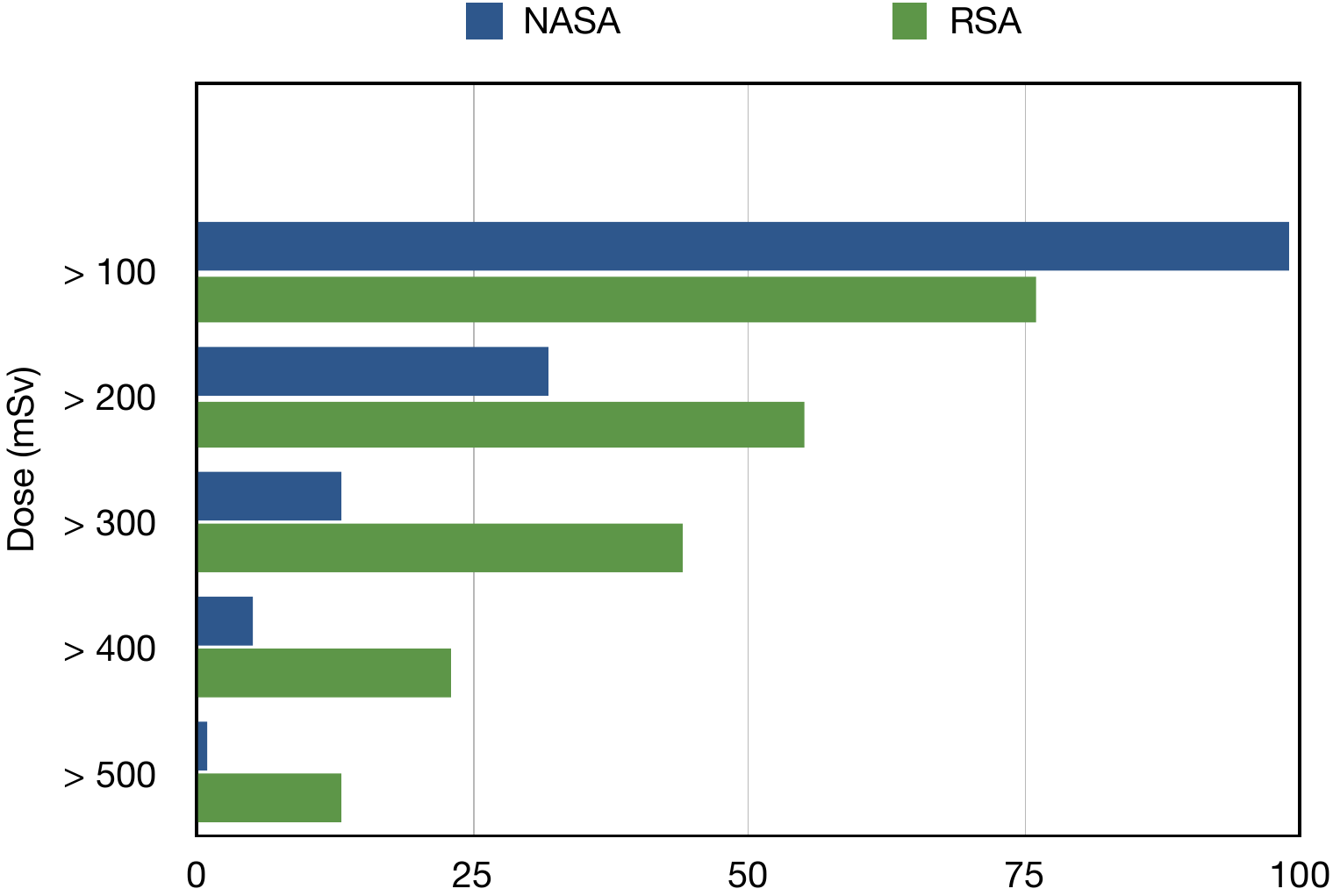}
\caption[]{{\bf Cumulative mission dose for all astronauts and cosmonauts through 2020.} Shown are the number of astronauts (NASA, ESA, CSA, JAXA) and cosmonauts whose cumulative mission exposures have exceeded dose thresholds up to 500 mSv. It should be noted that these were long-duration flyers, e.g., were missions of 3 months or longer are typical.}
\label{fig:fig4}
\end{figure*}

Since the onset of crewed spaceflight, it has been presumed that exposure to space radiation increases the risks of astronauts developing cancer, experiencing central nervous system decrements, exhibiting degenerative tissue effects or developing acute radiation syndrome \cite{carnell2016risk,chancellor2014space,mewaldt2005cosmic,board1997radiation,ncrp2006information,ncrp1988guidance,kennedy2014biological,boerma2015space,cucinotta2009risk,hagen1989radiation}. The majority of epidemiologic data results from the astronaut cohort are from exposures incurred on missions during the Space Shuttle era, where less than 100mSv was accumulated by an astronaut. Over the past decade, however, the nominal mission length for astronauts has increased to at least 6 months in duration with exposures of 1 mSv to 1.5 mSv per day, depending on the phase of the solar cycle, number of spacewalks performed, and the level of solar activity. Even with increasing mission length and radiation exposures (e.g. Fig.~\ref{fig:fig4}), it is noteworthy that to date no astronaut has been diagnosed with a cancer that is attributable to space radiation. Although the sample size is small, followup times for large exposures are limited, and cancer latency periods are years to decades. In addition, the neurocognitive deficits and vascular endothelial dysfunction leading to increased cardiovascular mortality has not been demonstrated compared to analog populations.

The limitations in understanding and data on radiogenic risks to astronauts comprise major obstacles to making meaningful and reliable , predictions of human clinical outcomes.  These also hamper efforts to develop and inform risk mitigation strategies before, during, or after exposure. The presumed risk of cancer, cardiovascular diseases and cognitive deficits due to space radiation remains one of NASA’s top limiting factors as we return to the moon and push forward towards Mars. The limitations in knowledge of radiogenic risks  been one of the key factors that can limit an astronaut’s potential flight assignments as well as  active spaceflight-career length.

Fig.~\ref{fig:fig4} demonstrates the amount of data obtained by both NASA and the Russian Space Agency (RSA) in regards to number of astronauts and dose of radiation obtained over varying lengths of spaceflight. Recently Zeitlin et al., reported findings that the LET spectrum measured inside ISS at high latitudes was very similar to measurements made both in lunar orbit and in interplanetary space during transit to Mars\cite{zeitlin2019comparisons}. This demonstrates that we now have an increasing data set of human population that has flown in space while exposed to doses that in some instances, exceed the identified thresholds for some degenerative and carcinogenic outcomes. Given the decades of ground-based radiobiologic research on radiation effects in space travel, one can only conclude that animal models and the type of radiation utilized in these experiments are not adequate  surrogates for the complex physiologic response of the human body to the complex radiation environment in space. Mice are profoundly different from humans and it remains very difficult to clinically translate the effects noted in the murine population to astronauts.

\section*{Wait, so astronauts can sign off on being launched into space, but not on an unproven health risk? }
For decades there have been concerns about the clinical sequelae of United States astronauts’ exposure to the complex radiation environment of spaceflight. The National Council of Radiation Protection and Measurements (NCRP) recommended to NASA that a career limit be placed on astronaut radiation exposure which when adjusted for age and sex, would not increase an astronaut’s risk of death by greater than 3\%. This risk of exposure-induced death (REID) was first suggested by NRCP Report 98 and is based on a two-sided 95\% confidence interval for estimates of mortality from malignancy \cite{ncrp1988guidance}. Little justification was given as to why a 3\% REID was chosen, relative to a higher or lower threshold. At that time, it was suggested that a 25 year old male astronaut would be able to accomplish 17 Space Shuttle missions of 90 days duration over a 10 year period prior before violating this REID. This proposed limit represented a reasonable ceiling given the available information as no astronaut in the 1980’s flew at this frequency or duration. However, the United States space program has fundamentally changed and astronauts now regularly complete 6 to 12 month missions aboard the International Space Station. Long-duration missions to Mars and the Moon are expected during the next two decades  and could last 1-3 years.

In clinical practice, medical providers and bioethicists often refer to four principles which guide medical decision making: autonomy, non-maleficence, beneficence, and justice. The first of these principles, autonomy, requires patients to have independent thought, intention and action when making decisions regarding their own health \cite{beauchamp2001principles}. In order to abide by the principle of autonomy, patients must fully understand the risks and benefits of any medical decision, including options which increase personal risk or include non-treatment. Patients must be empowered to have informed consent about the treatment options, or lack thereof, for any medical condition. Fortunately, in common clinical medical scenarios such as chest pain, pediatric fever, or the risk of pulmonary embolism, there are clinical decision rules based on robust, prospective, and externally validated clinical trials that stratify the patient’s individual risk.

Exploration of space is not without risks. Space agencies have rigorous safety programs and oversight to limit risk during missions planning, spacecraft design, and throughout the stages of flight. With regards to space radiation, it is important to acknowledge current limitations in the understanding of the impacts on human morbidity and mortality. The estimations of risk from space radiation are likely not as precise as the risk of vehicle loss during launch or re-entry. As noted in this manuscript, there are no validated clinical decision rules that medical providers and flight surgeons can use to precisely assess the clinical risk from space radiation. However, the principle of autonomy requires that medical providers provide patients, in this case astronauts, with the best information available and its limitations so they can make informed decisions regarding the risks from space radiation. Using such information, an astronaut may wish to exceed their strict personal 3\% REID for the opportunity to fly specific missions or achievements, such as returning to the Moon or being the first to travel to Mars.

Exploration of space, however, is not an individual endeavour. The launch and safe return of astronauts is supported by thousands of engineers, scientists, administrators, and the American public. Missions further science on behalf of citizens and taxpayers, not individual astronauts. Although we believe astronauts should be empowered to make decisions about their personal risk tolerance from radiation, such practices when left unregulated, may lead to group-think and development of unsafe practices. Astronauts and leaders at government and commercial spaceflight companies face a challenge of balancing individual autonomy, medical risk, and the proper stewardship of agency resources.

Given the nature of exploration spaceflight, extremely strict limits of radiation exposure may not be practical. Missions to the Moon or Mars may violate an astronaut’s 3\% REID simply due to mission duration. We believe that further research is needed to provide insight into an astronaut’s individual risk of morbidity and mortality from space radiation. This information is critical so that mutual, pragmatic, and ethically sound decision-making can take place between astronauts, medical providers, and space agency administrators
\section*{So what does all of this mean?}
The major current limitations of space radiation effects research are inhibited by limitations in the methods currently available to researchers. This includes a lack of suitable accelerator-producted radiation sources that has precluded the accurate reproduction of GCR at terrestrial laboratories.  A second major limitation is the lack of mammalian analogs that fully represent human physiology \cite{chancellor2014space}. Bench studies have used of mono-energetic beams and acute, single-ion exposures or multiple ions in sequence, instead of the poly-energetic spectra of multiple ion species present in the space radiation environment \cite{durante2005cytogenetic,durante2002karyotypes,wang2014heavier}. Furthermore, for ease of dose specification and modeling, ions in the 100 MeV/n to 1000 MeV/n range were often used, such that the entire target was contained within the plateau portion of the depth-dose distribution as demonstrated in Figure 3 \cite{chancellor2018limitations,kennedy2008effects,wambi2009protective}. Additionally, a projected, cumulative mission dose is often delivered to the animals over a single acute exposure; as such, the experimental dose rates are several orders of magnitude higher than actual space exposures \cite{chancellor2018limitations,wambi2009protective}. Even the more recently used prolonged fractionated exposures may not fully mimic the continuous low dose rates found in space.

The ability to better simulate the space radiation environment through terrestrial research efforts would greatly facilitate advances in our understanding of the risks to human health during long-duration spaceflight. There has been progress towards realistic ground-based space radiation analogs. NASA’s GCR simulator may be able to provide some improvements to simulation studies by use of rapid-sequential mono-electric beam exposures\cite{simonsen2020nasa}. There is debate in the space radiation community with respect to the appropriate order of ion exposures delivered, where the alteration of exposure sequence can affect the outcomes of an experiment\cite{chancellor2018limitations,kennedy2014biological,elmore2011neoplastic}. Sequential beam exposures remain ineffective in modeling complex and simultaneous exposures of the actual GCR environment\cite{norbury2016galactic,slaba2015gcr}. Additionally, Chancellor et al., have demonstrated that a moderator block can be placed in a single-ion-species beam without any requirement to modify the beamline and related infrastructure \cite{chancellor2017targeted}. As an iron beam passes through the moderator block, nuclear spallation processes create desired fragment spectra, resulting in fluences of charged particles  with atomic numbers of 1 $\leq$ Z $\leq$ 26 and LETs up to approximately 200 keV/$\mu$m. That is to say, the ion species and their distributions in energy closely mimic those of GCR. Taken together, the limitations of terrestrial GCR-like sources and limitation of animal models have been major obstacles to understanding responses at chronic, low-dose and low-dose- rate radiation conditions. Even partially eliminating one of these limitations may help researchers to  explain the frequent disparity between data from ground-based studies and those from astronauts.

\section*{Disclosure Statement}
The view(s) expressed herein are those of the authors and do not reflect the official policy or position of Louisiana State University, Brooke Army Medical Center, Mary Bird Perkins Cancer Center, the University of Rochester, Johnson Space Center, the National Aeronautics and Space Administration, the U.S. Army Medical Department, the U.S. Army Office of the Surgeon General, the Department of the Army, the Department of the Air Force, or the Department of Defense or the U.S. Government.

\section*{Acknowledgments}{\setlength\parindent{0pt} JCC acknowledges support by the Translational Research Institue for Space Health (TRISH) through NASA NNX16AO69A. JCC and MEC acknowledges support from LaSPACE/NASA Grant Number 80NSSC20M0110. JPW acknowledges support from the National Institute of Health through grants R01 CA220467 and R01 HL127001.
}

%%%%%%%%%%%%%%%%%%%%%%%%%%%%%%%%%%%%%%%%%%

\section*{Author Contributions}{All authors contributed to the content and reviewed the manuscript.}

%%%%%%%%%%%%%%%%%%%%%%%%%%%%%%%%%%%%%%%%%%


\begin{thebibliography}{10}
\expandafter\ifx\csname url\endcsname\relax
  \def\url#1{\texttt{#1}}\fi
\expandafter\ifx\csname urlprefix\endcsname\relax\def\urlprefix{URL }\fi
\providecommand{\bibinfo}[2]{#2}
\providecommand{\eprint}[2][]{\url{#2}}

\bibitem{carnell2016risk}
\bibinfo{author}{Carnell, L.}, \bibinfo{author}{Blattnig, S.},
  \bibinfo{author}{Hu, S.} \emph{et~al.}
\newblock \bibinfo{title}{Risk of acute radiation syndromes due to solar
  particle events: Evidence report} (\bibinfo{year}{2016}).

\bibitem{chancellor2018limitations}
\bibinfo{author}{Chancellor, J.~C.} \emph{et~al.}
\newblock \bibinfo{title}{Limitations in predicting the space radiation health
  risk for exploration astronauts}.
\newblock \emph{\bibinfo{journal}{npj Microgravity}}
  \textbf{\bibinfo{volume}{4}}, \bibinfo{pages}{1--11} (\bibinfo{year}{2018}).

\bibitem{chancellor2014space}
\bibinfo{author}{Chancellor, J.~C.}, \bibinfo{author}{Scott, G.~B.} \&
  \bibinfo{author}{Sutton, J.~P.}
\newblock \bibinfo{title}{Space radiation: the number one risk to astronaut
  health beyond low earth orbit}.
\newblock \emph{\bibinfo{journal}{Life}} \textbf{\bibinfo{volume}{4}},
  \bibinfo{pages}{491--510} (\bibinfo{year}{2014}).

\bibitem{cucinotta2006evaluating}
\bibinfo{author}{Cucinotta, F.~A.}, \bibinfo{author}{Kim, M.-H.~Y.} \&
  \bibinfo{author}{Ren, L.}
\newblock \bibinfo{title}{Evaluating shielding effectiveness for reducing space
  radiation cancer risks}.
\newblock \emph{\bibinfo{journal}{Radiation Measurements}}
  \textbf{\bibinfo{volume}{41}}, \bibinfo{pages}{1173--1185}
  (\bibinfo{year}{2006}).

\bibitem{guetersloh2006polyethylene}
\bibinfo{author}{Guetersloh, S.} \emph{et~al.}
\newblock \bibinfo{title}{Polyethylene as a radiation shielding standard in
  simulated cosmic-ray environments}.
\newblock \emph{\bibinfo{journal}{Nuclear Instruments and Methods in Physics
  Research Section B: Beam Interactions with Materials and Atoms}}
  \textbf{\bibinfo{volume}{252}}, \bibinfo{pages}{319--332}
  (\bibinfo{year}{2006}).

\bibitem{newhausershielding}
\bibinfo{author}{Newhauser, W.}, \bibinfo{author}{Dexheimer, D.} \&
  \bibinfo{author}{Titt, U.}
\newblock \bibinfo{title}{Shielding of proton therapy facilities: A review of
  recent progress} .

\bibitem{wilson1999shielding}
\bibinfo{author}{Wilson, J.~W.} \emph{et~al.}
\newblock \bibinfo{title}{Shielding from solar particle event exposures in deep
  space}.
\newblock \emph{\bibinfo{journal}{Radiation measurements}}
  \textbf{\bibinfo{volume}{30}}, \bibinfo{pages}{361--382}
  (\bibinfo{year}{1999}).

\bibitem{zeitlin2005shielding}
\bibinfo{author}{Zeitlin, C.}, \bibinfo{author}{Guetersloh, S.},
  \bibinfo{author}{Heilbronn, L.} \& \bibinfo{author}{Miller, J.}
\newblock \bibinfo{title}{Shielding and fragmentation studies}.
\newblock \emph{\bibinfo{journal}{Radiation protection dosimetry}}
  \textbf{\bibinfo{volume}{116}}, \bibinfo{pages}{123--124}
  (\bibinfo{year}{2005}).

\bibitem{slaba2017optimal}
\bibinfo{author}{Slaba, T.~C.} \emph{et~al.}
\newblock \bibinfo{title}{Optimal shielding thickness for galactic cosmic ray
  environments}.
\newblock \emph{\bibinfo{journal}{Life Sciences in space research}}
  \textbf{\bibinfo{volume}{12}}, \bibinfo{pages}{1--15} (\bibinfo{year}{2017}).

\bibitem{williams2010animal}
\bibinfo{author}{Williams, J.~P.} \emph{et~al.}
\newblock \bibinfo{title}{Animal models for medical countermeasures to
  radiation exposure}.
\newblock \emph{\bibinfo{journal}{Radiation research}}
  \textbf{\bibinfo{volume}{173}}, \bibinfo{pages}{557--578}
  (\bibinfo{year}{2010}).

\bibitem{williams2016addressing}
\bibinfo{author}{Williams, J.~P.} \emph{et~al.}
\newblock \bibinfo{title}{Addressing the symptoms or fixing the problem?
  developing countermeasures against normal tissue radiation injury}.
\newblock \emph{\bibinfo{journal}{Radiation research}}
  \textbf{\bibinfo{volume}{186}}, \bibinfo{pages}{1--16}
  (\bibinfo{year}{2016}).

\bibitem{chancellor2017targeted}
\bibinfo{author}{Chancellor, J.~C.} \emph{et~al.}
\newblock \bibinfo{title}{Targeted nuclear spallation from moderator block
  design for a ground-based space radiation analog}.
\newblock \emph{\bibinfo{journal}{arXiv preprint arXiv:1706.02727}}
  (\bibinfo{year}{2017}).

\bibitem{simpson1983elemental}
\bibinfo{author}{Simpson, J.}
\newblock \bibinfo{title}{Elemental and isotopic composition of the galactic
  cosmic rays}.
\newblock \emph{\bibinfo{journal}{Annual Review of Nuclear and Particle
  Sciences}} \textbf{\bibinfo{volume}{33}} (\bibinfo{year}{1983}).

\bibitem{mewaldt2005cosmic}
\bibinfo{author}{Mewaldt, R.}
\newblock \bibinfo{title}{The cosmic ray radiation dose in interplanetary space
  present day and worst-case evaluations}  (\bibinfo{year}{2005}).

\bibitem{rahmanifard2020galactic}
\bibinfo{author}{Rahmanifard, F.} \emph{et~al.}
\newblock \bibinfo{title}{Galactic cosmic radiation in the interplanetary space
  through a modern secular minimum}.
\newblock \emph{\bibinfo{journal}{Space Weather}}
  \textbf{\bibinfo{volume}{18}}, \bibinfo{pages}{e2019SW002428}
  (\bibinfo{year}{2020}).

\bibitem{wouter2020crater}
\bibinfo{author}{Wouter, C.} \emph{et~al.}
\newblock \bibinfo{title}{Crater observations and permissible mission duration
  for human operations in deep space}.
\newblock \emph{\bibinfo{journal}{Life Sciences in Space Research}}
  (\bibinfo{year}{2020}).

\bibitem{cucinotta2006cancer}
\bibinfo{author}{Cucinotta, F.~A.} \& \bibinfo{author}{Durante, M.}
\newblock \bibinfo{title}{Cancer risk from exposure to galactic cosmic rays:
  implications for space exploration by human beings}.
\newblock \emph{\bibinfo{journal}{The lancet oncology}}
  \textbf{\bibinfo{volume}{7}}, \bibinfo{pages}{431--435}
  (\bibinfo{year}{2006}).

\bibitem{board1997radiation}
\bibinfo{author}{Board, S.~S.}, \bibinfo{author}{Council, N.~R.} \emph{et~al.}
\newblock \emph{\bibinfo{title}{Radiation hazards to crews of interplanetary
  missions: biological issues and research strategies}}
  (\bibinfo{publisher}{National Academies Press}, \bibinfo{year}{1997}).

\bibitem{edwards2001rbe}
\bibinfo{author}{Edwards, A.}
\newblock \bibinfo{title}{Rbe of radiations in space and the implications for
  space travel.}
\newblock \emph{\bibinfo{journal}{Physica Medica: PM: an International Journal
  Devoted to the Applications of Physics to Medicine and Biology: Official
  Journal of the Italian Association of Biomedical Physics (AIFB)}}
  \textbf{\bibinfo{volume}{17}}, \bibinfo{pages}{147--152}
  (\bibinfo{year}{2001}).

\bibitem{ncrp2006information}
\bibinfo{author}{NCRP.}
\newblock \bibinfo{title}{Information needed to make radiation protection
  recommendations for space missions beyond low-earth orbit}
  (\bibinfo{organization}{National Council on Radiation Protection and
  Measurements}, \bibinfo{year}{2006}).

\bibitem{ncrp1988guidance}
\bibinfo{author}{NCRP.}
\newblock \emph{\bibinfo{title}{Guidance on radiation received in space
  activities}} (\bibinfo{publisher}{NCRP Report No. 98}, \bibinfo{year}{1988}).

\bibitem{hellweg2007getting}
\bibinfo{author}{Hellweg, C.~E.} \& \bibinfo{author}{Baumstark-Khan, C.}
\newblock \bibinfo{title}{Getting ready for the manned mission to mars: the
  astronauts’ risk from space radiation}.
\newblock \emph{\bibinfo{journal}{Naturwissenschaften}}
  \textbf{\bibinfo{volume}{94}}, \bibinfo{pages}{517--526}
  (\bibinfo{year}{2007}).

\bibitem{smart2003comment}
\bibinfo{author}{Smart, D.} \& \bibinfo{author}{Shea, M.}
\newblock \bibinfo{title}{Comment on estimating the solar proton environment
  that may affect mars missions}.
\newblock \emph{\bibinfo{journal}{Advances in Space Research}}
  \textbf{\bibinfo{volume}{31}}, \bibinfo{pages}{45--50}
  (\bibinfo{year}{2003}).

\bibitem{nelson2016space}
\bibinfo{author}{Nelson, G.~A.}
\newblock \bibinfo{title}{Space radiation and human exposures, a primer}.
\newblock \emph{\bibinfo{journal}{Radiation research}}
  \textbf{\bibinfo{volume}{185}}, \bibinfo{pages}{349--358}
  (\bibinfo{year}{2016}).

\bibitem{hackam2006translation}
\bibinfo{author}{Hackam, D.~G.} \& \bibinfo{author}{Redelmeier, D.~A.}
\newblock \bibinfo{title}{Translation of research evidence from animals to
  humans}.
\newblock \emph{\bibinfo{journal}{Jama}} \textbf{\bibinfo{volume}{296}},
  \bibinfo{pages}{1727--1732} (\bibinfo{year}{2006}).

\bibitem{perel2007comparison}
\bibinfo{author}{Perel, P.} \emph{et~al.}
\newblock \bibinfo{title}{Comparison of treatment effects between animal
  experiments and clinical trials: systematic review}.
\newblock \emph{\bibinfo{journal}{Bmj}} \textbf{\bibinfo{volume}{334}},
  \bibinfo{pages}{197} (\bibinfo{year}{2007}).

\bibitem{kennedy2014biological}
\bibinfo{author}{Kennedy, A.~R.}
\newblock \bibinfo{title}{Biological effects of space radiation and development
  of effective countermeasures}.
\newblock \emph{\bibinfo{journal}{Life sciences in space research}}
  \textbf{\bibinfo{volume}{1}}, \bibinfo{pages}{10--43} (\bibinfo{year}{2014}).

\bibitem{nolan2019emerging}
\bibinfo{author}{Nolan, M.~W.}, \bibinfo{author}{Kent, M.~S.} \&
  \bibinfo{author}{Keara~Boss, M.}
\newblock \bibinfo{title}{Emerging translational opportunities in comparative
  oncology with companion canine cancers: Radiation oncology}.
\newblock \emph{\bibinfo{journal}{Frontiers in Oncology}}
  \textbf{\bibinfo{volume}{9}}, \bibinfo{pages}{1291} (\bibinfo{year}{2019}).

\bibitem{siranart2016mixed}
\bibinfo{author}{Siranart, N.}, \bibinfo{author}{Blakely, E.~A.},
  \bibinfo{author}{Cheng, A.}, \bibinfo{author}{Handa, N.} \&
  \bibinfo{author}{Sachs, R.~K.}
\newblock \bibinfo{title}{Mixed beam murine harderian gland tumorigenesis:
  predicted dose-effect relationships if neither synergism nor antagonism
  occurs}.
\newblock \emph{\bibinfo{journal}{Radiation Research}}
  \textbf{\bibinfo{volume}{186}}, \bibinfo{pages}{577--591}
  (\bibinfo{year}{2016}).

\bibitem{edmondson2020genomic}
\bibinfo{author}{Edmondson, E.} \emph{et~al.}
\newblock \bibinfo{title}{Genomic mapping in outbred mice reveals overlap in
  genetic susceptibility for hze ion--and $\gamma$-ray--induced tumors}.
\newblock \emph{\bibinfo{journal}{Science Advances}}
  \textbf{\bibinfo{volume}{6}}, \bibinfo{pages}{eaax5940}
  (\bibinfo{year}{2020}).

\bibitem{walker2013heavy}
\bibinfo{author}{Walker, S.~A.}, \bibinfo{author}{Townsend, L.~W.} \&
  \bibinfo{author}{Norbury, J.~W.}
\newblock \bibinfo{title}{Heavy ion contributions to organ dose equivalent for
  the 1977 galactic cosmic ray spectrum}.
\newblock \emph{\bibinfo{journal}{Advances in Space Research}}
  \textbf{\bibinfo{volume}{51}}, \bibinfo{pages}{1792--1799}
  (\bibinfo{year}{2013}).

\bibitem{blue2019limitations}
\bibinfo{author}{Blue, R.~S.} \emph{et~al.}
\newblock \bibinfo{title}{Limitations in predicting radiation-induced
  pharmaceutical instability during long-duration spaceflight}.
\newblock \emph{\bibinfo{journal}{npj Microgravity}}
  \textbf{\bibinfo{volume}{5}}, \bibinfo{pages}{1--9} (\bibinfo{year}{2019}).

\bibitem{hall2006radiobiology}
\bibinfo{author}{Hall, E.~J.}, \bibinfo{author}{Giaccia, A.~J.} \emph{et~al.}
\newblock \emph{\bibinfo{title}{Radiobiology for the Radiologist}},
  vol.~\bibinfo{volume}{6} (\bibinfo{publisher}{Lippincott Williams \&
  Wilkins}, \bibinfo{year}{2006}).

\bibitem{icrp1998icrp}
\bibinfo{author}{ICRP}.
\newblock \emph{\bibinfo{title}{ICRP Publication 78: Individual Monitoring for
  Internal Exposure of Workers}}.
\newblock \bibinfo{number}{78} (\bibinfo{publisher}{Elsevier Health Sciences},
  \bibinfo{year}{1998}).

\bibitem{attix2008introduction}
\bibinfo{author}{Attix, F.~H.}
\newblock \emph{\bibinfo{title}{Introduction to radiological physics and
  radiation dosimetry}} (\bibinfo{publisher}{John Wiley \& Sons},
  \bibinfo{year}{2008}).

\bibitem{dasu2013impact}
\bibinfo{author}{Dasu, A.} \& \bibinfo{author}{Toma-Dasu, I.}
\newblock \bibinfo{title}{Impact of variable rbe on proton fractionation}.
\newblock \emph{\bibinfo{journal}{Medical physics}}
  \textbf{\bibinfo{volume}{40}}, \bibinfo{pages}{011705}
  (\bibinfo{year}{2013}).

\bibitem{boerma2015space}
\bibinfo{author}{Boerma, M.} \emph{et~al.}
\newblock \bibinfo{title}{Space radiation and cardiovascular disease risk}.
\newblock \emph{\bibinfo{journal}{World journal of cardiology}}
  \textbf{\bibinfo{volume}{7}}, \bibinfo{pages}{882} (\bibinfo{year}{2015}).

\bibitem{cucinotta2009risk}
\bibinfo{author}{Cucinotta, F.~A.} \& \bibinfo{author}{Durante, M.}
\newblock \bibinfo{title}{Risk of radiation carcinogenesis}.
\newblock \emph{\bibinfo{journal}{Human health and performance risks of space
  exploration missions. NASA SP-2009-3405. Houston: National Aeronautics and
  Space Administration}} \bibinfo{pages}{119--170} (\bibinfo{year}{2009}).

\bibitem{hagen1989radiation}
\bibinfo{author}{Hagen, U.}
\newblock \bibinfo{title}{Radiation biology in space: a critical review}.
\newblock \emph{\bibinfo{journal}{Advances in Space Research}}
  \textbf{\bibinfo{volume}{9}}, \bibinfo{pages}{3--8} (\bibinfo{year}{1989}).

\bibitem{zeitlin2019comparisons}
\bibinfo{author}{Zeitlin, C.} \emph{et~al.}
\newblock \bibinfo{title}{Comparisons of high-linear energy transfer spectra on
  the iss and in deep space}.
\newblock \emph{\bibinfo{journal}{Space Weather}}
  \textbf{\bibinfo{volume}{17}}, \bibinfo{pages}{396--418}
  (\bibinfo{year}{2019}).

\bibitem{beauchamp2001principles}
\bibinfo{author}{Beauchamp, T.~L.}, \bibinfo{author}{Childress, J.~F.}
  \emph{et~al.}
\newblock \emph{\bibinfo{title}{Principles of biomedical ethics}}
  (\bibinfo{publisher}{Oxford University Press, USA}, \bibinfo{year}{2001}).

\bibitem{durante2005cytogenetic}
\bibinfo{author}{Durante, M.} \emph{et~al.}
\newblock \bibinfo{title}{Cytogenetic effects of high-energy iron ions:
  dependence on shielding thickness and material}.
\newblock \emph{\bibinfo{journal}{Radiation research}}
  \textbf{\bibinfo{volume}{164}}, \bibinfo{pages}{571--576}
  (\bibinfo{year}{2005}).

\bibitem{durante2002karyotypes}
\bibinfo{author}{Durante, M.}, \bibinfo{author}{George, K.},
  \bibinfo{author}{Wu, H.} \& \bibinfo{author}{Cucinotta, F.}
\newblock \bibinfo{title}{Karyotypes of human lymphocytes exposed to
  high-energy iron ions}.
\newblock \emph{\bibinfo{journal}{Radiation research}}
  \textbf{\bibinfo{volume}{158}}, \bibinfo{pages}{581--590}
  (\bibinfo{year}{2002}).

\bibitem{wang2014heavier}
\bibinfo{author}{Wang, H.} \& \bibinfo{author}{Wang, Y.}
\newblock \bibinfo{title}{Heavier ions with a different linear energy transfer
  spectrum kill more cells due to similar interference with the ku-dependent
  dna repair pathway}.
\newblock \emph{\bibinfo{journal}{Radiation research}}
  \textbf{\bibinfo{volume}{182}}, \bibinfo{pages}{458--461}
  (\bibinfo{year}{2014}).

\bibitem{kennedy2008effects}
\bibinfo{author}{Kennedy, A.~R.}, \bibinfo{author}{Davis, J.~G.},
  \bibinfo{author}{Carlton, W.} \& \bibinfo{author}{Ware, J.~H.}
\newblock \bibinfo{title}{Effects of dietary antioxidant supplementation on the
  development of malignant lymphoma and other neoplastic lesions in mice
  exposed to proton or iron-ion radiation}.
\newblock \emph{\bibinfo{journal}{Radiation research}}
  \textbf{\bibinfo{volume}{169}}, \bibinfo{pages}{615--625}
  (\bibinfo{year}{2008}).

\bibitem{wambi2009protective}
\bibinfo{author}{Wambi, C.~O.} \emph{et~al.}
\newblock \bibinfo{title}{Protective effects of dietary antioxidants on proton
  total-body irradiation-mediated hematopoietic cell and animal survival}.
\newblock \emph{\bibinfo{journal}{Radiation research}}
  \textbf{\bibinfo{volume}{172}}, \bibinfo{pages}{175--186}
  (\bibinfo{year}{2009}).

\bibitem{simonsen2020nasa}
\bibinfo{author}{Simonsen, L.~C.}, \bibinfo{author}{Slaba, T.~C.},
  \bibinfo{author}{Guida, P.} \& \bibinfo{author}{Rusek, A.}
\newblock \bibinfo{title}{Nasa’s first ground-based galactic cosmic ray
  simulator: Enabling a new era in space radiobiology research}.
\newblock \emph{\bibinfo{journal}{PLoS biology}} \textbf{\bibinfo{volume}{18}},
  \bibinfo{pages}{e3000669} (\bibinfo{year}{2020}).

\bibitem{elmore2011neoplastic}
\bibinfo{author}{Elmore, E.}, \bibinfo{author}{Lao, X.},
  \bibinfo{author}{Kapadia, R.}, \bibinfo{author}{Swete, M.} \&
  \bibinfo{author}{Redpath, J.}
\newblock \bibinfo{title}{Neoplastic transformation in vitro by mixed beams of
  high-energy iron ions and protons}.
\newblock \emph{\bibinfo{journal}{Radiation research}}
  \textbf{\bibinfo{volume}{176}}, \bibinfo{pages}{291--302}
  (\bibinfo{year}{2011}).

\bibitem{norbury2016galactic}
\bibinfo{author}{Norbury, J.~W.} \emph{et~al.}
\newblock \bibinfo{title}{Galactic cosmic ray simulation at the nasa space
  radiation laboratory}.
\newblock \emph{\bibinfo{journal}{Life sciences in space research}}
  \textbf{\bibinfo{volume}{8}}, \bibinfo{pages}{38--51} (\bibinfo{year}{2016}).

\bibitem{slaba2015gcr}
\bibinfo{author}{Slaba, T.~C.} \emph{et~al.}
\newblock \bibinfo{title}{Gcr simulator reference field and a spectral approach
  for laboratory simulation}.
\newblock \emph{\bibinfo{journal}{NASA Technical Paper}}
  \textbf{\bibinfo{volume}{218698}}, \bibinfo{pages}{2015}
  (\bibinfo{year}{2015}).

\end{thebibliography}
\end{document}